\documentclass[]{kluwer}
\usepackage{epsf}
\usepackage[dvips]{epsfig}

\begin{document}

\begin{article}

\begin{opening}         
\title{$\omega$ Cen - an Ultra Compact Dwarf Galaxy ?} 
\author{Michael \surname{Fellhauer}$^{1,2}$, Pavel \surname{Kroupa}$^{2}$}  
\runningauthor{M. Fellhauer, P. Kroupa}
\runningtitle{$\omega$ Cen - an Ultra Compact Dwarf Galaxy ?}
\institute{$^{1}$ School of Mathematics, Univ.\ Edinburgh, Scotland,
  U.K.\\ 
  $^{2}$ Inst.\ Theor.\ Phys.\ \& Astrophys., Univ.\ Kiel, Germany} 
\date{\today}

\begin{abstract}
  We study the merging of star clusters out of cluster
  aggregates similar to Knot S in the Antennae on orbits close to
  the one of $\omega$ Cen by carrying out high resolution numerical
  N-body simulations. We want to constrain the parameter space
  which is able to produce merger objects with similar properties
  as $\omega$ Cen.
\end{abstract}

\keywords{globular clusters: individual: $\omega$ Cen -- methods:
  N-body simulations -- galaxies: formation -- galaxies: star clusters
  -- galaxies: dwarfs} 
\end{opening}           

\section{Introduction}
\label{sec:intro}

Interactions of gas-rich disk galaxies show intense bursts of star
formation.  For example HST-images of the Antennae (Whitmore et al.\
1999) reveal that the knots of intense star formation produce clusters
of massive young star clusters.  These aggregates which we call
super-clusters ($=$ cluster of star clusters; not to confuse with
super stellar cluster (SSC), which are individual massive star
clusters) appear to contain dozens to hundreds of massive star
clusters within a region spanning only a few hundred pc to a kpc in
radius. 

On the other hand new observations of the central galaxy of the
Fornax cluster revealed a new class of unresolved, compact
objects (Hilker et al.\ 1999, Phillipps et al.\ 2001) with radii
of a few hundred pc, which are called ultra compact dwarf galaxies
(UCD).  Also in a few lenticular field galaxies (e.g.\ NGC~1023)
Larson and Brodie (2000, 2002) found star clusters with extremely
large effective radii ($r_{\rm eff} > 7$~pc) which they call faint
fuzzies.  

Here we show N-body results concerning the dynamical evolution of such
super-cluster aggregates.  All simulations of super-clusters show a
strong merging behaviour building up compact merger objects in few
super-cluster crossing times (Fellhauer et al.\ 2002).  Depending on
the initial conditions of our simulations (strong or weak tidal field;
massive or extended low-mass super-cluster) our resulting merger
objects have similar properties like the new classes of objects above
(Fellhauer \& Kroupa 2002a/b). 

But placing compact and massive super-clusters in strong tidal fields
on an orbit similar to $\omega$-Cen reveals an object which has
similar properties like the most massive globular cluster (GC) in the
Milky Way. $\omega$-Cen is not only the most massive GC, it has also
some strange properties like different populations of stars (different
ages and metalicities).  It shows signs of rotation with a maximum
rotation speed of $8$~km/s (Freeman 2001).

\section{Setup}
\label{sec:setup}

We use the particle-mesh code {\sc Superbox} (Fellhauer et al.\ 2000)
which incorporates a hierarchical grid architecture allowing high
resolution at the places of interest. 

We model the single star clusters as Plummer spheres (Plummer 1911,
numerical realisation: Aarseth et al.\ 1974) 
with a Plummer radius of
$4$~pc, which corresponds to the mean half-light radius found for the
individual young star clusters in the Antennae (Whitmore et al.\
1999).  $N_{0} = 262$ star clusters with a mass  range of $10^{4} -
10^{6}$~M$_{\odot}$ following a power-law mass spectrum,  
%%\begin{eqnarray}
%%  \label{eq:massspec}
 $ n(M_{\rm cl})  \propto  M_{\rm cl}^{-1.5}$, 
%%\end{eqnarray}
are placed in a Plummer distribution with Plummer radius $r_{\rm
  pl}^{\rm sc}$ of $20$~pc, a cut-off radius of $100$~pc and a total
mass of $M_{\rm sc}= 10^{7}$~M$_{\odot}$ representing the
super-cluster.  The crossing time of the super-cluster is $t_{\rm
  cr}^{\rm sc} =   2.6$~Myr and the velocity dispersion of the
clusters in the super-cluster is $\sigma_{\rm sc} =
25.2$~kms$^{-1}$.  Additionally we choose the sense of rotation of all
clusters in the super-cluster to be the same to investigate the
resulting rotation-law of the merger object.  A super-cluster is
expected to rotate if it forms from a contracting and locally
differentially rotating inner tidal arm.  

The super-cluster is placed on an eccentric orbit with perigalacticon
at $2.1$~kpc and apogalacticon at $7.5$~kpc.  The orbit is inclined
such that the maximum $Z$-distance from the disc plane is about
$2$~kpc.  The parameters are chosen to be representative of the knots
seen to contain many star clusters in the Antennae galaxies, while the
orbital inclination is motivated by the orbit of $\omega$-Cen (Dinescu 
et al.\ 1999).  The host galaxy is represented by an analytical
potential, which consists of a disc modelled as a Plummer-Kuzmin
potential and a spherical  halo component modelled as a logarithmic
potential:   
\begin{eqnarray}
  \label{eq:pot}
  \Phi_{\rm gal} & = & \Phi_{\rm disc} + \Phi_{\rm halo} \\
   & = & - \frac{GM_{\rm disc}}{\sqrt{R'^{2} + ( a +
      \sqrt{Z^{2}+b^{2}})^{2} } } - \frac{1}{2} v_{0}^{2} \ln(
  R_{\rm gal}^{2} + R^{2}), \nonumber
\end{eqnarray}
with $M_{\rm disc} = 10^{11}$~M$_{\odot}$, $a = 3$~kpc, $b = 0.3$~kpc,
$v_{0} = 200$~km/s and $R_{\rm gal} = 50$~kpc which sums up to an
almost flat rotation curve with a rotation speed of $220$~kms$^{-1}$. 

It is possible to follow the evolution with a particle-mesh code that
neglects dynamical effects of two-body relaxation, because the
half-mass (bulk) two-body relaxation time of the single star clusters,
which can be estimated from (Binney \& Tremaine 1987) 
\begin{eqnarray}
  \label{eq:relax}
  t_{\rm relax} & = & \frac{664}{\ln(0.5N)} \left( \frac{M_{\rm
        cl}} {10^{5}{\rm M}_{\odot}} \right)^{1/2} \left(
    \frac{1{\rm M}_{\odot}} {\overline{m}} \right) \left(
    \frac{r_{0.5}} {1 {\rm pc}} \right)^{3/2} {\rm Myr},
\end{eqnarray}
is $\approx 800$~Myr for a $10^{4}$~M$_{\odot}$ star cluster ranging
up to $4.4$~Gyr for a $10^{6}$~M$_{\odot}$ star cluster, while the
merging timescale is much shorter (Fellhauer et al.\ 2002).
Furthermore, as shown below, the resulting merger objects have
relaxation times of a Hubble-time or longer.   

\section{Results}
\label{sec:res}

After the merging process is over ($\approx 150$~Myr) the object loses
mass due to tidal 
shaping on its eccentric orbit.  Because of the collision-free code no
mass-loss due to internal evolution (evaporation because of two-body
encounters $=$ two-body relaxation) is taken into account.  But taking
Eq.~\ref{eq:relax} the merger object has a relaxation time of about
60~Gyr, therefore mass-loss due to evaporation should be a minor
effect.  The bound mass of the merger object is about $8.5 \cdot
10^{6}$~M$_{\odot}$ after formation and after $10$~Gyr of tidal
shaping it still has $4.5 \cdot 10^{6}$~M$_{\odot}$.  The mass after
formation is smaller than the sum of the merged star clusters because
of massloss during the violent merging process.  
Most of this unbound material gets spread along the orbit of
the merger object but some of these stars can still be found in the
neighbourhood or even within the merger object.  One can see this
clearly in the line-of-sight velocity dispersion.  This causes a rise
in velocity dispersion  shortly within and beyond the tidal radius,
which is due to these unbound stars which have a totally different
velocity signature than the bound stars. 

After $10$~Gyr the half-mass radius of the object is $19$~pc and the
total size (tidal radius) is about $100$~pc (93 at perigalacticon and
120 at apogalacticon).  Fitting a King-profile to the surface density
distribution gives a core radius ($=$ effective or half-light radius)
of $8.5$~pc and a central surface density of
$9500$~M$_{\odot}/$pc$^{2}$ which corresponds to a central surface
brightness (taking $M/L=3.0$) of about $18$~mag./arcsec$^{2}$.  A
better fit to the data would be an exponential profile in the inner
part with a power law profile with power index $-5.6$ for the outer
part.  The 3D velocity dispersion of the merger object is about
24~km/s, while the line-of-sight velocity dispersion is about 14~km/s.
The maximum rotation velocity of our object is 4~km/s.  
\begin{table}[h!]
  \caption{Properties of $\omega$-Cen.}
  \label{tab:ocen}
  \begin{center}
    \begin{tabular}[h!]{ll}
      galactocentric distance   & $6.7$~kpc \\
      total luminosity          & $M_{V}=-10.3$~Mag. \\
      total mass ($M/L=4.1$)    & $5.1 \cdot 10^{6}$~M$_{\odot}$
      \\ 
      core radius               & $3.7$~pc \\
      half-mass radius          & $6.1$~pc \\
      tidal radius              & $64.6$~pc \\
      velocity dispersion       & $21.9$~km/s \\
      maximum rotation velocity & $8$~km/s \\
      metalicity                & -1.62 (mean); -1.8 to -0.8 \\
      age                       & 15 Gyr; age spread 4 Gyr
    \end{tabular}
  \end{center}
\end{table}

\section{Outlook}
\label{sec:out}

Although our best model does not yet have exactly the same properties
as $\omega$-Cen, we think that we are on the right way to solve the
puzzle of the origin of $\omega$-Cen.  Our merger object is still not
heavy enough and too large compared to $\omega$-Cen.

A possible explanation of the age and metalicity spread and the high
rotation speed could be an underlying population of old stars stemming
from a dissolved dwarf galaxy plunging into the Milky Way and causing
the star burst and the formation of the super-cluster, which leads to
the building of $\omega$-Cen in the above described scenario.
This scenario is being studied with numerical experiments that have
started recently. \\

\end{article}

\end{document}